# Fast Incremental and Personalized PageRank

Bahman Bahmani[*]
Stanford University
bahman@stanford.edu

Abdur Chowdhury
Twitter Inc.
abdur@twitter.com

Ashish Goel[†]
Twitter Inc. and Stanford University
ashishg@stanford.edu

## ABSTRACT

In this paper, we analyze the efficiency of Monte Carlo methods for incremental computation of PageRank, personalized PageRank, and similar random walk based methods (with focus on SALSA), on large-scale dynamically evolving social networks. We assume that the graph of friendships is stored in distributed shared memory, as is the case for large social networks such as Twitter.

For global PageRank, we assume that the social network has $n$ nodes, and $m$ adversarially chosen edges arrive in a random order. We show that with a reset probability of $\epsilon$, the total work needed to maintain an accurate estimate (using the Monte Carlo method) of the PageRank of every node at all times is $O(\frac{n \ln m}{\epsilon^2})$. This is significantly better than all known bounds for incremental PageRank. For instance, if we naively recompute the PageRanks as each edge arrives, the simple power iteration method needs $\Omega(\frac{m^2}{\ln(1/(1-\epsilon))})$ total time and the Monte Carlo method needs $O(mn/\epsilon)$ total time; both are prohibitively expensive. Furthermore, we also show that we can handle deletions equally efficiently.

We then study the computation of the top $k$ personalized PageRanks starting from a seed node, assuming that personalized PageRanks follow a power-law with exponent $\alpha < 1$. We show that if we store $R > q \ln n$ random walks starting from every node for large enough constant $q$ (using the approach outlined for global PageRank), then the expected number of calls made to the distributed social network database is $O(k/(R^{(1-\alpha)/\alpha}))$.

We also present experimental results from the social networking site, Twitter, verifying our assumptions and analyses. The overall result is that this algorithm is fast enough for real-time queries over a dynamic social network.

## 1. INTRODUCTION

[*]Work done while interning at Twitter

[†]Work done while at Twitter. Ashish Goel's Research was also supported by NSF grant IIS-0904325.

Over the last decade, PageRank [39] has emerged as a very effective measure of reputation for both web graphs and social networks (where it was historically known as eigenvector centrality [15]). Also, collaborative filtering has proved to be a very effective method for personalized recommendation systems [10, 32, 37]. In this paper, we will focus on fast incremental computation of (approximate) PageRank, personalized PageRank [14, 19, 39], and similar random walk based methods, particularly SALSA [30] and personalized SALSA [38, 40], over dynamic social networks, and its applications to reputation and recommendation systems over these networks. Incremental computation is useful when edges in a graph arrive over time, and it is desirable to update the PageRank values right away rather than wait for a batched computation.

Surprisingly, despite the fact that computing PageRank is a well studied problem [5], some simple assumptions on the structure of the network and the data layout lead to dramatic improvements in running time, using the simple Monte Carlo estimation technique.

In large scale web applications, the underlying graph is typically stored on disk, and either edges are streamed [11] or a map-reduce computation is performed. The Monte Carlo method requires random access to the graph, and has not found widespread practical use in these applications.

However, for social networking applications, it is crucial to support random access to the underlying network, since messages flow on edges in a network in real-time, and random access to the social graph is necessary for the core functionalities of the network. Hence, the graph is usually stored in distributed shared memory, which we denote as "Social Store", providing a data access model very similar to Scalable Hyperlink Store [36]. We use this feature strongly in obtaining our results.

In this introduction, we will first provide some background on PageRank and SALSA and typical approaches to computing them, and then outline our results along with some basic efficiency comparisons. The literature related to the problem studied in this paper is really vast, and in the final subsection of this introduction, we do a comprehensive review of this literature, and compare our results with the related previous results.

### 1.1 Background

In this paper, we focus on (incremental) computation of PageRank [39], personalized PageRank [14, 19, 39], SALSA [30], and personalized SALSA [38, 40]. So, in this subsection, we provide a quick review of these methods. Here and throughout the paper, we denote the number of nodes in the network by $n$, and the number of edges in the network by $m$.

PageRank is the stationary distribution of a random walk which, at each step, with a certain probability $\epsilon$ jumps to a random node, and with probability $1 - \epsilon$ follows a randomly chosen outgoing edge from the current node. Personalized PageRank is the same as PageRank, except that all the jumps are made to the seed node for which we are personalizing the PageRanks.

SALSA, just like HITS [23], associates two scores with each node $v$, called hub score, $h_v$, and authority score $a_v$. These scores represent how good a hub or authority each node is. The scores are related as follows:

$$h_v = \sum_{\{x|\ (v,x) \in E\}} a_x/\text{indeg}(x)$$

$$a_x = \sum_{\{v|\ (v,x) \in E\}} h_v/\text{outdeg}(v)$$

where $E$ is the set of edges of the graph, and indeg($x$) and outdeg($v$) are, respectively, the in-degrees and out-degrees of the nodes. Notice that SALSA corresponds to a forward-backward random walk, where the walk alternates between forward and backward steps. The personalized version of SALSA, that we consider, allows random jumps (to the seed node) at forward steps. Thus, personalizing over node $u$ corresponds to the following equations:

$$h_v = \epsilon \delta_{u,v} + (1-\epsilon) \sum_{\{x|\ (v,x) \in E\}} a_x/\text{indeg}(x)$$

$$a_x = \sum_{\{v|\ (v,x) \in E\}} h_v/\text{outdeg}(v)$$

Notice that in our setting, hub scores and authority scores can be interpreted, respectively, as similarity measures and relevance measures. Hence, we obtain a simple system for recommending additional friends to a user: just recommend those with the highest relevance. We will now outline two broad approaches to computing PageRank and SALSA; a more detailed overview is presented in section 1.3. The first approach is to use linear algebraic techniques, the simplest of which is the power iteration method. In this method, the PageRank of node $v$ is initialized to $\pi_0(v) = 1/n$ and the following update is repeatedly performed:

$$\forall v, \pi_{i+1}(v) = \epsilon/n + \sum_{\{w|\ (w,v) \in E\}} \pi_i(w)(1-\epsilon)/\text{outdeg}(w). \quad (1)$$

This gives exponential reduction in error per iteration, resulting in a running time of $O(m)$ per iteration and $O(m/\ln(1/(1-\epsilon)))$ for getting the total error down to a constant. The other broad approach is Monte Carlo, where we directly do random walks to estimate the PageRank of each node. In the simplest instantiation, we do $R$ "short random walks" of geometric lengths (with mean $1/\epsilon$) starting at each node. Each short random walk simulates one continuous session by a random surfer who is doing the PageRank random walk. While the error does not decay as fast as in the power iteration method, $R = \ln(n/\epsilon)$ or even $R = 1$ give provably good results, and have running time of $O(nR/\epsilon)$, which, for non-sparse graphs, is much better than that of power iteration. SALSA can be computed using obvious modifications to either approach.

## 1.2 Our Results

We study two problems. First, efficient incremental computation of PageRank (and its variants) over dynamically evolving networks. Second, efficiently computing personalized PageRank (and its variants) under the power-law network model. Both of these problems are considered to be among the most important problems regarding PageRank [26]. Below, we overview our results in more detail.

We present formal analyses of incremental computation of random walk based reputations and collaborative filters in the context of evolving social networks. In particular, we focus on very efficiently approximating both global and personalized variants of PageRank [39] and SALSA [30]. We perform experiments to validate each of the assumptions made in our analysis, and also study the empirical performance of our algorithms.

For global PageRank, we show that in a network with $n$ nodes, and $m$ adversarially chosen edges arriving in random order, we can maintain very accurate estimates of the PageRank (and authority score) of every node at all times with only $O(\frac{n \ln m}{\epsilon^2})$ total work. This is a dramatic improvement over the naïve running time $\Omega(\frac{m^2}{\ln(1/(1-\epsilon))})$ (e.g., by recomputing the PageRanks upon arrival of each edge, using the power iteration method) or $\Omega(\frac{mn}{\epsilon})$ (e.g., using the Monte Carlo method from scratch each time an edge arrives). Similarly, we show that in a network with $m$ edges, upon removal of a random edge, we can update all the PageRank approximations using only $O(n/m\epsilon^2)$ expected work. Our algorithm is a Monte Carlo method [3] that works by maintaining a small number of short random walk segments starting at each node in the social graph. The same approach works for SALSA. For global SALSA, the authority score of a node is exactly its in-degree as the reset probability goes to 0, so the primary reason to store these random walk segments is to aid in computing personalized SALSA scores. It is important to note that it is only the efficiency of our algorithms that depends on the random order assumption; it is also important to note that the random order assumption is weaker than assuming the well-known generative models for power-law networks, such as Preferential Attachment [4].

We then study the problem of finding the $k$ nodes with highest personalized PageRank values (or personalized authority scores). We show that we can use the same building blocks used for global PageRank and SALSA, that is, the stored walk segments at each node, to very efficiently find very accurate approximations for the top $k$ nodes. We prove that, assuming that the personalized scores follow a power-law with exponent $0 < \alpha < 1$, if we cache $R > q \ln n$ random walk segments starting at every node (for large enough constant $q$), then the expected number of calls made to the distributed social network database is $O(k/R^{\frac{1-\alpha}{\alpha}})$. This is significantly better than $n$ and even $k$. Notice that without the power-law assumption, in general, one would have to find the scores of all the nodes and then find the top $k$ results (hence, one would need at least $\Omega(n)$ work).

We present the results of the experiments we did to validate our assumptions and analyses. We used the network data from the social networking site Twitter. The access to this data was through a database, called FlockDB, stored in distributed shared memory. Our experiments support our random order assumption on edge arrivals (or at least the specific claim that we need for our results to go through). Also, we observe that not only do the global PageRank scores and in-degrees follow the same power-laws (as previously proved under mild assumptions in the literature [33]), but also the personalized PageRanks follow power-laws with average exponent roughly equal to the exponent for PageRank and in-degree. Finally, our experiments also support the proved theoretical bounds on the number of calls to the social network database.

Random walk based methods have been reported to be very effective for the link prediction problem on social networks [31]. We also did some preliminary experiments to explore this further. The results are presented in appendix A and indicate that random walk based algorithms (i.e., personalized PageRank and personalized SALSA) significantly outperform HITS as a recommendation system for twitter users; we present this comparison not as significant original research but as an interesting data point for readers who are interested in practical aspects of recommendation systems.

### 1.3 Related Work

Any PageRank computation or approximation method on social networks is desired to have the following properties:

1. Ability to keep the values (or approximations) updated all the time as the network evolves

2. Large scale full personalization capability

3. Very high computational efficiency

Also, as briefly mentioned above, in social networking applications, the data access model is dictated by the need for random access to the network, and implemented using a Social Store. Therefore, a desirable PageRank computation (or approximation) scheme should achieve the above mentioned features in this data access model.

A simple way to keep the PageRank values updated is to just recompute the values for each incremental change in the network. But, this can be very costly. For instance, the simple power iteration method [39] to approximate (to a constant precision) PageRank values (with reset probability $\epsilon$) takes $\Omega(\frac{x}{\ln(1/(1-\epsilon))})$ time over a graph with $x$ edges. Hence, over $m$ edge arrivals, this takes $\sum_{x=1}^{m} \Omega(\frac{x}{\ln(1/(1-\epsilon))})$ $= \Omega(\frac{m^2}{\ln(1/(1-\epsilon))})$ total time, which is many orders of magnitude larger than our approach. Similarly, the $\Omega(n/\epsilon)$ time complexity of the Monte Carlo method results in a total $\Omega(mn/\epsilon)$ work over $m$ edge arrivals, which is also very inefficient. So, we need more efficient ways of doing the computations.

There have been a lot of methods proposed for computation or approximation of PageRank and similar measures [5]. Broadly speaking, these methods can be categorized into two general categories, based on the core techniques they use:

1. Linear algebraic methods: These methods mainly use techniques from linear and matrix algebra, perhaps with some application of structural properties of the networks of interest (e.g., the world wide web) [8,9,16, 19–22, 24, 25, 27–29, 35, 41–43].

2. Monte Carlo methods: These methods use a small number of simulated random walks per node to approximate PageRanks (or other variants) [3,11,13,41].

A great survey of many of the methods in the first category is done by Langville and Meyer [26]. However, for completeness and also to compare the state of the art with our own results, we provide an overview of the methods and results in this category here.

A family of the methods proposed in this category deal with accelerating the basic power iteration method for computing PageRank values [20–22]. However, they all provide only very modest (i.e., small constant factor) speed ups. For instance, Kamvar et al. [22] propose a method to accelerate the power iteration, using an extrapolation based on Aitken $\Delta^2$ method for accelerating linearly convergent sequences. However, as discussed in their paper, the time complexity of their method is $\Omega(n)$, which is prohibitively large for a real-time application. Also, their experiments show only a $25 - 300\%$ speed up compared to the crude power iteration. So, the method does not perform well enough for our applications.

Another family of the methods in the first category deal with efficiently updating the PageRank values using the "aggregation" idea [9,24,25,27–29]. The basic idea behind these methods is that when an incremental change happens in the network, the effect of this change on the PageRank vector is mostly local. That is, only the PageRanks of the nodes in the vicinity of the change may change significantly. To utilize this observation, these methods partition the set of the network nodes to two subsets $G, \bar{G}$, where $G$ is a subset of nodes close to where the incremental change happened, and $\bar{G}$ is the set of all other nodes. Then, all the nodes in $\bar{G}$ are lumped/aggregated into a single hyper-node, so a smaller network (composed of $G$ and this hyper-node) is formed. Then, the PageRanks of the nodes in $G$ are updated using this network, and finally the result is translated back to the original network.

None of these methods seem well suited for real time applications. First, the performance of these methods heavily depends on the partitioning of the network, and as pointed out in [27], a bad choice of this partitioning can cause these methods to be as slow as the power iteration. It is not known how to do this partitioning; while a number of heuristic ideas have been proposed [9, 25], there is also considerable evidence that these networks are expanders, and no such partitioning is possible. Further, it is easy to see that independent of how the partitioning is done, partitioning and aggregation together will need $\Omega(n)$ time. Also, notice that this work is in addition to the actual PageRank computation that needs to be done on the aggregated network, and this computational load is also not negligible. For instance, as reported by Chien et al. [9], for a single edge addition to a network with $60M$ nodes, they need to do a PageRank computation on a network with almost $8K$ nodes. After all, these methods start with a precise PageRank vector and give an approximate PageRank vector for the network after the incremental change. Therefore, even if these methods were run for a real-time application, the approximation error would potentially accumulate, and the estimations would drift away from the true PageRanks [24]. Of course, we should mention that there exist exact aggregation based update methods, but all of those methods are more costly than power iteration [25]!

A number of other methods in the first category also deal with updating PageRanks [35, 43]. However, the method in [43] does not scale well for the large scale social networking applications. It achieves $O(l^2)$ update time for random walk based methods on an $n \times l$ bipartite graph. This may work well when the graph is very skewed (i.e., $l << n$). But, for instance, in the friend recommendation application on social networks, $l = n$, so this gives only $O(n^2)$ update time, and hence $O(mn^2)$ total time over $m$ edge arrivals, which is very bad. Also, McSherry [35] combines a number of heuristics to provide some improvement in computation and updating of PageRank, using a sequential implementation of the power iteration. The method works in the streaming model where edges are stored and then streamed from the disk. No guarantee is given about the tradeoff between precision and the time complexity of the method.

Another family of methods in the first category [16, 19, 21] deals with personalization. In this family, Haveliwala's work [16] achieves personalization only to the level of few (e.g., 16) topics and provides no efficiency improvement over the power iterations.

Kamvar et al. [21] use the host-induced block structure of the web link graph to speed up computation and updating of PageRank and also provide personalization. However, first, in social networks, there is no equivalent of a web host, and more generally it is not easy to find a corresponding block structure (even if such a structure actually exists). Therefore, it is not even clear how to apply this idea to social networking applications. Also, they use the block structure only to come up with a better initialization for a standard PageRank computation, such as power iteration. Therefore, even though, after an incremental change, the local PageRank vectors (corresponding to unchanged hosts) may be reused, doing the power iteration alone would need $\Omega(m)$ work per each change (and hence a total $\Omega(m^2)$ work over $m$ edge arrivals). Finally, on the personalization front, their method achieves personalization only at the host level.

Jeh and Widom [19] achieve personalization only over a subset of the network nodes, denoted as the "Hub Set". Even though, the paper provides no precise time complexity analysis, it is easy to see that, as mentioned in the paper, the time performance of the presented algorithm heavily depends on the choice of the hub set. In our application where we wish to have full personalization, the hub set needs to be simply the set of all vertices, in which case the algorithms in [19] reduce to a simple dynamic programming which provides no performance improvement over power iteration.

Another notable work in the first category is [41]. It uses deterministic rounding or randomized sketching techniques along with the dynamic programming approach proposed in [19] to achieve full personalization. However, the time complexity of their (rounding) method, to achieve constant error, is $O(m/\epsilon)$, while, the time complexity of the simple Monte Carlo method to achieve constant error with high probability is just $O(n \ln n/\epsilon)$. Therefore, if $m = \omega(n \ln n)$, which is expected in our applications of interest, then the simple Monte Carlo method is asymptotically faster than the method introduced in [41]. Also, it is not clear how would one be able to efficiently update this method's estimations as the network changes.

The methods in the second category above, namely Monte Carlo methods, have the advantage that they are very efficient and can achieve full personalization on a large scale [3, 13]. However, all the literature in this category deals with static networks. Of course, it has been mentioned in [3] that one can keep the approximations updated continuously. However, they neither provide any details of how exactly to do this nor give any analysis about the efficiency of doing these updates. For instance, the method in [3] uses $\Omega(\frac{n}{\epsilon})$ work in each computation. So, if we naively recompute the PageRank using this method for each edge arrival, then over $m$ edge arrivals, we will have $\Omega(\frac{mn}{\epsilon})$ total work.

In contrast, in this paper, we show that one can use the Monte Carlo techniques to achieve very cheap incremental updates. Indeed, we prove a surprising result, stating that up to a logarithmic factor, the total work required to keep the approximations updated all the time is the same as the work needed to just initialize the approximations! More precisely, we prove that over $m$ edge arrivals in a random order, we can keep the approximations updated using only $O(\frac{n \ln m}{\epsilon^2})$ total work. This is significantly better than all the previously proposed methods for doing the updates.

Another issue with the methods in the second category is that if we want to directly use the simulated random walk segments to approximate personalized PageRank, we would get a limited precision. For instance, if we store $R$ random walks per node (and use only the end points of these walks for our estimates) the approximate personalized PageRank vectors that we get would have at most $R$ non-zero entries, which is significantly fewer than what we need in all applications of interest; previous works [11, 13] do not explore this tradeoff in any detail.

In this paper, we present a formal analysis of this trade off in the random access model, and prove that under the power-law model for the network, one can do long enough walks (and hence get desirable levels of precision) very efficiently. Indeed, Das Sarma et al. [11] achieve time performance $O(m/\sqrt{\epsilon})$, and hence using their method for each incremental change in the network would need $O(m^2/\sqrt{\epsilon})$ total work over $m$ edge arrivals. While, this is better than the naive power iteration $O(m^2/\ln(1/(1-\epsilon)))$ time complexity, it is still very inefficient. But, we show that in our model, we can achieve an $O(n \ln m/\epsilon^2)$ time complexity, assuming a power-law network model, which is significantly better, and good enough for real-time incremental applications.

Collaborative filtering on social networks is very closely related to the Link Prediction problem [1, 12, 17, 18, 31, 42], for which the random walk based methods have been reported to work very well [31] (we also verified this experimentally; the results are given in Appendix A). Most of the literature on Link Prediction deals with static networks. But, there are some proposed algorithms which deal with dynamic evolving networks [1, 43]. However, none of these methods scale to today's very large scale social networks.

We would also like to mention that there are incremental collaborative filtering methods based on low-rank SVD updates. For instance, refer to [7] and references therein. However, these methods also do not scale very well. For instance, the method proposed in [7] requires a total $O(pqr)$ work to calculate the rank-$r$ SVD of a $p \times q$ matrix. But, for instance, for the friend recommendation application, $p = q = n$, and hence this method needs a total $\Omega(n^2)$ work, while we only need a total $O(\frac{n \ln m}{\epsilon^2})$ work, as mentioned above, which is significantly better.

## 2. INCREMENTAL COMPUTATION OF PAGERANK

In this section, we explain the details of the Monte Carlo method for approximating the (global) PageRank values. Then, we will prove our surprising result on the total amount of work needed to keep the PageRank estimates updated all the time. Even though we will focus on PageRank, our results will extend (with some minor modifications that will be pointed out) to other random walk based methods, such as SALSA.

### 2.1 Approximating PageRank

To approximate PageRank, we do $R$ random walks starting at each node of the network. Each of these random walks is continued until its first reset (Hence, each one has average length $1/\epsilon$). We store all these walk segments in a database, where each segment is stored at every node that it passes through. Assume for each node $v$, $X_v$ is the total number of times that any of the stored walk segments visit $v$. Then, we approximate the PageRank of $v$, denoted by $\pi_v$, with:

$$\widetilde{\pi}_v = \frac{X_v}{nR/\epsilon}$$

Then, we have the following theorem:

THEOREM 1. $\widetilde{\pi}_v$ is sharply concentrated around its expectation, which is $\pi_v$.

This theorem is proved in Appendix B. The exact concentration bounds follow from the proof. But, to summarize, the obtained approximations are quite good even for $R = 1$. Thus, the defined $\widetilde{\pi}_v$'s are accurate approximations of the actual PageRank values $\pi_v$.

### 2.2 Updating the Approximations

As the underlying network evolves, the PageRank values of the nodes also change. For a lot of applications, we would like to have updated approximations of the PageRank values all the time. Thus, here we analyze the cost of keeping the approximations updated at all times. To keep the approximations updated, we only need to keep the random walk segments stored at the network nodes updated. Thus, we analyze the amount of work in keeping these walks updated.

We first prove the following proposition:

PROPOSITION 2. Assume $(u_t, v_t)$ is the random edge arriving at time $t$ ($1 \leq t \leq m$). Define $M_t$ to be the number of random walk segments that need to be updated at time $t$ ($1 \leq t \leq m$). Then, we have:

$$E[M_t] \leq \frac{nR}{\epsilon} E[\frac{\pi_{u_t}}{outdeg_{u_t}(t)}]$$

where the expectation on the right hand side is over the edge arrival at time $t$ (i.e., $E[\cdot] = E_{(u_t, v_t)}[\cdot]$), and $outdeg_u(t)$ is the outdegree of node $u$ after $t$ edges have arrived.

PROOF. The basic intuition in this proposition is that most random walk segments miss most network edges. More precisely, a walk segment needs to change only if it passes through the node $u_t$ and the random step at $u_t$ picks $v_t$ as the next node in the walk. In expectation, the number of times each walk segment visits $u_t$ is $\frac{\pi_{u_t}}{\epsilon}$. For each such visit, the probability for the walk to need a reroute is $\frac{1}{outdeg_{u_t}(t)}$. Hence, by a union bound, the probability that a walk segment needs an update is at most $\frac{\pi_{u_t}}{\epsilon} \frac{1}{outdeg_{u_t}(t)}$. Also, there is a total of $nR$ walk segments. Therefore, by linearity of expectation:

$$E[M_t] \leq \sum_u \frac{nR}{\epsilon} \pi_u \frac{1}{outdeg_u(t)} Pr[u_t = u]$$

$$= \frac{nR}{\epsilon} E[\frac{\pi_{u_t}}{outdeg_{u_t}(t)}]$$

which proves the lemma. □

In the above proposition, $E[\frac{\pi_{u_t}}{outdeg_{u_t}(t)}]$ depends on the exact network growth model. The model that we assume here is the random permutation model, in which $m$ adversarially chosen directed edges arrive in random order. Notice that this is a weaker assumption on the network evolution model than most of the popular models, such as the Preferential Attachment model [4]. We also experimentally validate this model, and present the results, confirming that this is actually a very good assumption, later in this paper.

For this model, we have the following lemma:

LEMMA 3. If $(u_t, v_t)$ is the edge arriving at time $t$ ($1 \leq t \leq m$) in a random permutation over the edge set, then:

$$E[\frac{\pi_{u_t}}{outdeg_{u_t}(t)}] = \frac{1}{t}$$

PROOF. For random arrivals, we have:

$$Pr[u_t = u] = \frac{outdeg_u(t)}{t}$$

Hence,

$$E[\frac{\pi_{u_t}}{outdeg_{u_t}(t)}] = \sum_u \pi_u \frac{1}{outdeg_u(t)} Pr[u_t = u]$$

$$= \sum_u \pi_u \frac{1}{outdeg_u(t)} \frac{outdeg_u(t)}{t}$$

$$= \sum_u \frac{\pi_u}{t} = \frac{1}{t} \sum_u \pi_u = \frac{1}{t}$$

□

From lemma 3 and proposition 2, we get the following theorem:

THEOREM 4. The expected amount of update work as the $t^{th}$ network edge arrives is at most $nR/t\epsilon^2$, and the expected total amount of work needed to keep the approximations updated over $m$ edge arrivals is at most $\frac{nR}{\epsilon^2} \ln m$.

PROOF. Defining $M_t$ as in proposition 2, we know from the same proposition that

$$E[M_t] \leq \frac{nR}{\epsilon} E[\frac{\pi_{u_t}}{outdeg_{u_t}(t)}]$$

Also, from lemma 3, we know:

$$E[\frac{\pi_{u_t}}{outdeg_{u_t}(t)}] = \frac{1}{t}$$

Hence,
$$E[M_t] \leq \frac{nR}{\epsilon} \frac{1}{t}$$

For each walk segment that needs an update, we can redo the walk starting at the updated node, or even more simply starting at the corresponding source node. So, for each such walk segment, in average we need at most $1/\epsilon$ work (equal to the average length of the walk segment). Hence, the average work needed at time $t$ is at most $\frac{nR}{\epsilon^2}\frac{1}{t}$, as stated in the theorem.

Summing up over all time instances, we get that the expected total amount of work that we need to do over $m$ edge arrivals (to keep the approximations updated all the time) is:
$$\frac{nR}{\epsilon^2} \sum_{t=1}^{m} \frac{1}{t} = \frac{nR}{\epsilon^2} H_m \leq \frac{nR}{\epsilon^2} \ln m$$

where $H_m$ is the $m^{th}$ harmonic number. Therefore, the total amount of expected work required is at most $\frac{nR}{\epsilon^2} \ln m$, which finishes the proof. □

The above theorem bounds the amount of update work as new edges arrive. Similarly, we can show that we can very efficiently handle edges leaving the graph:

PROPOSITION 5. *When the network has $m$ edges, if a randomly chosen edge leaves the graph, then the expected amount of work necessary to update the walk segments is at most $nR/m\epsilon^2$.*

PROOF. If $M$ is the number of walk segments that need to be updated, and $(u^*, v^*)$ is the random edge leaving the network, then exactly as in Proposition 2, one can see that: $E[M] \leq \frac{nR}{\epsilon} E[\frac{\pi_{u^*}}{\text{outdeg}_{u^*}}]$, and exactly as in Lemma 3, one can see $E[\frac{\pi_{u^*}}{\text{outdeg}_{u^*}}] = 1/m$. Finally, as in Theorem 4, the proof is completed by noticing that for each walk segment needing an update, the expected amount of work is $1/\epsilon$. □

The result in Theorem 4 (and similarly Proposition 5) is quite surprising (at least to the authors). Hence, we will now discuss various aspects of it. First, notice that the amount of work needed to keep the approximations updated all the time is only logarithmically larger than the cost to initialize the approximations (i.e., $nR/\epsilon$). Also, it is clear that the marginal update cost for not-so-early edges (say edges after the $\Omega(n)$ first ones) is so small, that we can do the updates in real time even per social interaction (e.g., clicks, etc.)

Also, notice that in applications in networks such as the web graph, we can not do the real time updates. That is because there is no way to figure out the changes in the network, other than recrawling the network which is not feasible in real time. Also, random access to edges in the network is expensive. However, in social networking applications, the network and all the events happening on it are always available and visible to the network provider. Therefore, it is indeed possible to do the real time updates, and hence this method is very well suited for social networking applications.

It should be mentioned that the update cost analyzed above is the extra cost due to updating the PageRank approximations. In other words, as a new edge is added to the network it should be added to the database containing the network. We can keep the random walk segments in another database, say PageRank Store. For each node $v$, we also keep two counters: one, denoted by $W(v)$, keeping track of the number of walk segments visiting $v$, and one, denoted by $d(v)$, keeping track of the outdegree of $v$. Then, when a new edge arrives at node $v$, first we add it to the Social Store. Then, with probability $1 - (1 - 1/d(v))^{W(v)}$ we call the PageRank Store to do the updates, in which case the PageRank Store will incur an additional cost as analyzed above in Theorem 4. We are assuming that the preprocessing (to generate the random number) can be done for free, which is reasonable, as it does not require any extra network transmissions or disk accesses; without this assumption, the running time would be $O(m + \frac{n \ln m}{\epsilon^2})$, which is still much better than the existing results.

In theorem 4, we analyzed the update costs under the random permutation model. Another model of interest is the Dirichlet model, in which $Pr[u_t = u] = [d_u(t-1) + 1]/[t-1+n]$. Following the same proof steps as in theorem 4, we can again prove that the total expected update cost over $m$ edge arrivals in this model is $\frac{nR}{\epsilon^2} \ln(\frac{m+n}{n})$. Again, the total updates cost is only logarithmically growing, and the marginal update costs are small enough to allow real time updates. This also raises the question that does the same result hold in the adversarial arrival model? Interestingly, the answer to this question is negative. We show this with an example.

EXAMPLE 1. *Consider a network formed by a directed N-cycle $v_1, v_2, \ldots, v_N$, a node $u$, $N$ nodes $x_1, x_2, \ldots, x_N$, and $N$ nodes $y_1, y_2, \ldots, y_N$ (hence, the total number of nodes in the network is $n = 3N + 1$). Assume every $v_j$ ($1 \leq j \leq N$) has an edge to $u$; $u$ has an edge to every $x_j$ ($1 \leq j \leq N$); every $x_j$ ($1 \leq j \leq N$) has an edge to $u$; $v_1$ has an edge to every $y_j$ ($1 \leq j \leq N$), and every $y_j$ ($1 \leq j \leq N$) has an edge to $v_1$. Then, in this network, adding just one edge from $u$ to $v_1$ will force $\Omega(n)$ random walk segments to need to be updated. So, it is not true that in an adversarial edge arrival model, the amount of work needed at each time instance vanishes over time. In other words, the above-presented results indeed use the randomness in the arrival order of the edges.*

## 2.3 Extension to SALSA

To approximate the hub and authority scores in SALSA, we need to keep $2R$ random walk segments per node; $R$ random walks starting with a forward step from the seed node, and $R$ walks starting with a backward step. Then, the approximations are done similar to the case of PageRank, and the sharp concentration of the approximations can be proved in a similar way.

For the update cost, we notice that if $(u_t, v_t)$ is the edge arriving at time $t$, then, unlike the PageRank case where only $u_t$ could cause updates, both $u_t$ and $v_t$ can cause walk segments to need an update. Again, we assume the random permutation model for edge arrivals. Then:

$$Pr[u_t = u] = \frac{\text{outdeg}_u(t)}{t}$$

$$Pr[v_t = v] = \frac{\text{indeg}_v(t)}{t}$$

and we get the following theorem:

THEOREM 6. *The expected amount of work needed to keep the approximations updated over $m$ edge arrivals is at most $\frac{16nR}{\epsilon^2} \ln m$.*

Rather than presenting the complete proof, which follows exactly the same steps as the one for theorem 4, we just explain from where the extra factor 16 is appearing: Instead of $R$ walks we are storing $2R$ walks per node, introducing a factor of 2. Rather than $1/\epsilon$, each walk segment has average length $2/\epsilon$ (because we only allow resets at forward steps), introducing a factor of 4 (as $\epsilon$ appears in the bound as $\epsilon^2$). Also, each time an edge $(u_t, v_t)$ arrives, both $u_t$ and $v_t$ can cause updates, hence twice as many walks need to be updated at each time. These three modifications, together cause a factor 16 difference.

## 3. APPROXIMATING PERSONALIZED PAGERANK AND SALSA

In the previous section, we showed how we can approximate PageRank and SALSA by keeping a small number of random walk segments per each node. In this section, we show how we can reuse the same stored walk segments to also approximate the personalized variants of PageRank and SALSA. Again, we will focus the discussion on (personalized) PageRank. All the results also extend to the case of SALSA.

The idea of using simulated random walk segments to approximate personalized PageRank was introduced in [13]. However, in that paper, the personalized PageRank values are simply approximated by the frequency of node visits in the walk segments. This limits the precision of the approximations. To improve the precision, the paper proposes to stitch the walk segments together to form longer walks. However, they provide neither any details of how to do this nor any analysis of the trade off between the precision and the query time.

Das Sarma et al. [11] propose a method to stitch the walk segments to form longer walks and analyze its performance in the streaming model in which the graph is accessed by streaming from disk. In this paper, we use an almost identical algorithm to do the personalized PageRank approximations. However, since we work in the random access model, we will need a different analysis of the algorithm.

We start by explaining the algorithm. Here, the basic idea is that we will perform the personalized PageRank random walk, but opportunistically use the $R$ stored random walk segments (described in section 2) for each node, where possible. To access these walk segments, we have to query the database containing them. A query to this database for a node $u$ returns all $R$ walk segments starting at $u$ as well as all the neighbors of $u$. We call such a query a "fetch" operation. Then, taking a random walk starting from a source node $w$, based on the stored walk segments, can be done as presented in Algorithm 1.

The main cost in this algorithm is the fetch operations it does. Everything else is done in main memory which is very fast. Thus, we would like to analyze the number of fetches made by this algorithm.

But, unlike the case of global PageRank and SALSA, we notice that in applications of personalized PageRank and SALSA, what we are interested in is the nodes with the largest values of the personalized score. For instance, in a recommendation system based on personalized SALSA or

---

**Algorithm 1** Personalized PageRank Walk Using Walk Segments

**Input:** Source node $w$, required length $L$ of the walk
**Output:** A personalized PageRank walk $P_w$ for source node $w$ of length at least $L$

Start the walk at $w$: $P_w \leftarrow [w]$
**while** length($P_w$) < $L$ **do**
  $u \leftarrow$ last node in $P_w$
  Generate a uniformly random number $\beta \in [0, 1]$
  **if** $\beta < \epsilon$ **then**
    Reset the walk to $w$: $P_w \leftarrow P_w \cdot \text{append}(w)$
  **else**
    **if** $u$ has an unused walk segment $Q$ remaining in memory **then**
      Add $Q$ to the end of $P_w$: $P_w \leftarrow P_w \cdot \text{append}(Q)$
      Then, reset the walk to $w$: $P_w \leftarrow P_w \cdot \text{append}(w)$
    **else**
      **if** $u$ was previously fetched **then**
        Take a random edge $(u, v)$ out of $u$
        Add $v$ to the end of $P_w$: $P_w \leftarrow P_w \cdot \text{append}(v)$
      **else**
        Do a fetch at $u$
      **end if**
    **end if**
  **end if**
**end while**

---

personalized PageRank, we are only interested in the nodes with the largest authority scores, because the system is eventually going to find and recommend only those nodes anyway. Thus, our objective here is to find the $k$ nodes (for some suitably chosen $k$) with the largest personalized scores. We show that, under a power-law network model, the above algorithm does this very efficiently.

We start with first exactly explaining our network model.

### 3.1 Network Model

If $\overrightarrow{\pi}$ is the vector of the scores of interest (e.g., personalized PageRanks), we assume that $\overrightarrow{\pi}$ follows a power-law model. That is, if $\pi_j$ is the $j^{th}$ largest entry in $\overrightarrow{\pi}$, then we assume:

$$\pi_j \propto j^{-\alpha} \qquad (2)$$

for some $0 < \alpha < 1$. This is a very well-motivated model. First, it has been proved [33] that (under some assumptions), in a network with power-law indegrees, the PageRank vector also follows a power-law, which has the same exponent as the one for indegrees. Our experiments with the Twitter social network, whose results are presented in section 4, not only confirm this result, but also show that personalized PageRank vectors also follow power-laws, and that the average exponent for the personalized PageRank vectors is roughly the same as (yes, you guessed it!) the one for indegree and global PageRank.

By approximating summation with integration, we can approximately calculate the normalizing factor in equation 2, denoted by $\eta$:

$$1 = \sum_{j=1}^{n} \pi_j = \eta \sum_{j=1}^{n} j^{-\alpha} \simeq \eta n^{\alpha-1} \int_0^1 x^{-\alpha} dx = \eta n^{\alpha-1} \cdot \frac{1}{1-\alpha}$$

Hence $\eta = (1-\alpha)/n^{1-\alpha}$, and:

$$\pi_j = \frac{(1-\alpha)j^{-\alpha}}{n^{1-\alpha}} \quad (3)$$

So, we will assume that the values of $\pi_j$'s are given by equation 3 (i.e., we ignore the very small error in estimating the summation with integration). This completes the description of our model.

### 3.2 Approximating the top $k$ nodes

Fixing a number $c$, we do a long enough random walk (with resets to the seed node) that for each of the top $k$ nodes, we expect to see that node at least $c$ times. Then, we will return the $k$ nodes most visited in this random walk. To this end, we first give a definition and prove a small technical lemma.

DEFINITION 1. *$X_{s,v}$ is the number of times that we visit node $v$ in a random walk of length $s$.*

LEMMA 7. $\sum_v |E[X_{s,v}] - s.\pi_v| \leq 2/\epsilon$

The proof of this lemma is given in Appendix C.

The above lemma shows that if $s\pi_v$ is not very small (e.g., compared to $1/\epsilon$), we can approximate $E[X_{s,v}]$ (and because of sharp concentration of $X_{s,v}$, even $X_{s,v}$ itself) with $s\pi_v$. This is what we will do in the rest of this section.

Therefore, in order to see each of the top $k$ nodes $c$ times in expectation, the minimum length of the walk that we need to take is determined by $s\pi_k = c$, which gives:

$$s_k = \frac{c}{1-\alpha} k . \left(\frac{n}{k}\right)^{1-\alpha} \quad (4)$$

This gives us the length of the walk that we need to do using our algorithm. So, now we can analyze the algorithm. We prove the following theorem:

THEOREM 8. *If we store $R > q \ln n$ walk segments at each node for a large enough constant $q$, then the expected number of fetches done to take a random walk of length $s$ is at most: $1 + (2(1-\alpha)/nR)^{\frac{1}{\alpha}-1}.s^{1/\alpha}$*

PROOF. A fetch is made at node $v$ only if we arrive at $u$ from a parent node $v$ which ran out of unused walk segments. In other words, each fetch at a node $u$ can be charged to an extra visit to one of $u$'s parents. Therefore, denoting the number of fetches made during the algorithm by $F$, we have:

$$F \leq \sum_v (X_{s,v} - R)^+$$

Hence,

$$E[F] \leq \sum_v E[(X_{s,v} - R)^+]$$

$$= \sum_v E[X_{s,v} - R | X_{s,v} \geq R] Pr(X_{s,v} \geq R)$$

$$\leq n \sum_{\{v|\, E[X_{s,v}] < R/2\}} Pr(X_{s,v} \geq R) + \sum_{\{v|\, E[X_{s,v}] \geq R/2\}} E[X_{s,v}]$$

$$\leq 1 + \sum_{\{v|\, s\pi_v > R/2\}} E[X_{s,v}]$$

Where the second to last inequality holds because (due to the memoryless property of the random walk) $E[X_{s,v} - R | X_{s,v} \geq R] \leq E[X_{s,v}]$, and the last inequality holds because with $R > q \ln n$ for large enough $q$, if $E[X_{s,v}] < R/2$ then $Pr(X_{s,v} > R) = o(1/n)$ using Chernoff bounds (and if $E[X_{s,v}] \geq R/2$ then $E[X_{s,v}]$ is almost equal to $s\pi_v$, as mentioned after lemma 7).

But, $s\pi_v > R/2$ if and only if $v \leq \tau$ where $\tau$ is such that $s\pi_\tau = R/2$. This gives:

$$\tau = \frac{(1-\alpha)^{1/\alpha}}{n^{\frac{1}{\alpha}-1}} \left(\frac{2s}{R}\right)^{1/\alpha}$$

and $E[F] \leq 1 + \sum_{j=1}^\tau s\pi_j$. Upperbounding the summation with integration, we get:

$$E[F] \leq 1 + (1-\alpha)\int_0^{\tau/N} x^{-\alpha} dx = 1 + s(\tau/n)^{1-\alpha}$$

$$= 1 + \left(\frac{2(1-\alpha)}{nR}\right)^{\frac{1}{\alpha}-1}.s^{1/\alpha}$$

which finishes the proof. □

REMARK 1. *We defined a fetch operation to return all the stored walk segments as well as all the outgoing edges of the queried node. While the number of stored walks per each node is small, the outdegree of a node can be very large (indeed as large as $\Omega(n)$). For instance, in the Twitter social network, @BarackObama has more than $750,000$ outgoing edges. Thus, a fetch at such nodes may cause memory problems. However, notice that if we change the fetch operation for node $w$ to either return all $R$ stored walk segments starting at $w$ or just one randomly sampled outgoing edge from $w$, then in the analysis in theorem 8, we will only get at most a factor $2$ more fetches (because, we will have $F \leq 2\sum_v (X_{s,v} - R)^+$). So, we will just stick with our original definition of a fetch operation.*

Using the value of $s_k$ from equation 4 in the result of theorem 8, directly gives the following corollary:

COROLLARY 9. *Storing $R > q \ln n$ walk segments per node for a large enough constant $q$, the expected number of fetches needed to find the top $k$ nodes is at most $1 + \frac{c^{1/\alpha}}{(1-\alpha)(R/2)^{\frac{1}{\alpha}-1}} k$.*

It should be noted that the bounds given in theorem 8 and corollary 9 are, respectively, $O\left(\frac{s}{(nR/s)^{\frac{1-\alpha}{\alpha}}}\right)$ and $O\left(\frac{k}{R^{\frac{1-\alpha}{\alpha}}}\right)$. Also, as our experiments (described later) show, the theoretical bounds are fairly accurate for values of $R$ as small as 5.

REMARK 2. *To compare the bounds from equation 4 and corollary 9, let $\alpha = 0.75$, $c = 5$, $R = 10$, $k = 100$, and $n = 10^8$. Then, 4 bounds the number of required steps (also equal to number of database queries, if done in the crude way) with $632k = 63200$, while 9 bounds the number of required fetches with the much smaller number $20k = 2000$. Also, notice how significantly smaller than $n = 10^8$ both these bounds are. This is because we are taking advantage of the power-law assumption/property for the random walks we are interested in. Without this assumption, in general, even to find the top $k$ nodes, one would need to calculate all $n$ entries of the stationary distribution, and then return the top-k values.*

## 4. EXPERIMENTS

In this section, we present the results of the experiments that we did to test our assumptions and methods.

### 4.1 Experimental Setup

We used data from the social networking site, Twitter, for our experiments. This network consists of directed edges. The access to data was through Twitter's Social Store, called FlockDB, stored in distributed shared memory. We emulated the PageRank Store on top of FlockDB. We used the reset probability $\epsilon = 0.2$ in our experiments. For the personalized experiments, we picked 100 random users from the network, who had a reasonable number of friends (between 20 and 30).

### 4.2 Verification of the Random Permutation Assumption

Given a single permutation (i.e., the one that we actually observed on twitter), it is impossible to validate whether edges do arive in random order. However, we can validate some associated statistics, and as it turns out, these will also provide a sufficient precondition for our analysis to hold:

1. Let $X$ denote the expected value of $\pi_v/\text{outdeg}_v$ for an arriving edge $(v, w)$. We assumed that $mX = 1$ in our proof; this is the only assumption that we need in order for our proof to be applicable. In order to validate this assumption, we looked at 4.63 Million edges arriving between two snapshots of the social graph (we removed edges originating from new nodes). The average observed value of $mX$ for these 4.63 Million arrivals was 0.81. This validates the running time we obtained in this paper (in fact, this is a little better than what we assumed since smaller values of $mX$ imply reduced computation time).

2. While not strictly required for our proof, another interesting consequence of the random permutation model is that the probability of an edge arriving out of node $v$ is proportional to the out-degree of $v^1$. We will call this the proportionality assumption. Let $a(d)$ denote the fraction of newly arriving edges $(v, w)$ such that $\text{outdeg}_v \leq d$. We will refer to $a(d)$ as the arrival degree cdf (cumulative distribution function). Further, let $s(d)$ denote the sum of the degrees of all the nodes which have degree at most $d$, and let $e(d)$ denote $s(d)/m$. We will refer to $e(d)$ as the existing degree cdf. If the proportionality assumption is true, we would expect the two cdfs to nearly coincide. Figure 1 shows plots of these two cdfs; it is clear that the two cdfs indeed track each other quite well.

### 4.3 Network Model Verification

As we explained in section 3.1, we assumed a power-law model on the personalized PageRank values. In addition to considerable literature [33], this assumption was also based on our experiments, showing the following results:

1. Our network has power-law indegrees and global PageRank, with exponent roughly 0.76. The results of the experiment for this result is presented in figure 2.

---
[1]In fact, 1+ the out degree, but that distinction will not be empirically important.

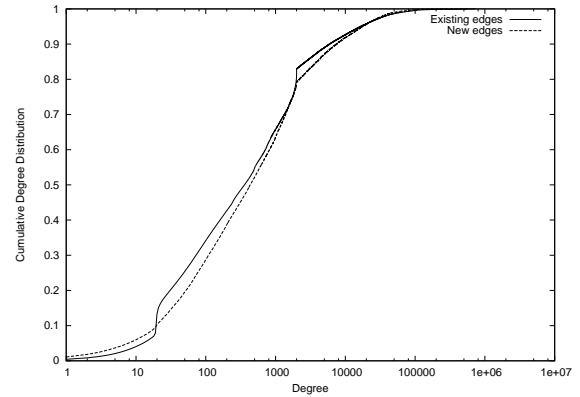

Figure 1: Arrival degree and existing degree cumulative distribution functions

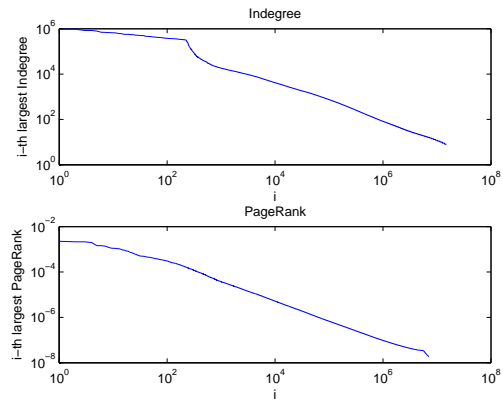

Figure 2: Indegree and PageRank power-laws. Both axes are log-scale.

2. Personalized PageRank vectors follow power-laws. The results for the personalized PageRank vectors of 6 random users are presented in figure 3. Around 2% of the nodes had $\alpha > 1$. Our analysis is easily adapted to this case, but we omit the details.

3. As shown in figure 4, there is a variation in the exponents of the power-laws followed by personalized PageRank vectors of different users. However, the mean of these exponents is almost the same as the exponent for indegree and PageRank. In our experiment, the average exponent was 0.77 and the standard deviation was 0.08.

REMARK 3. *The number on top of each plot in figure 3 shows the number of friends of the corresponding user. Notice that there is an initial part in each of these plots which follows a different power-law than the bulk of the vector. This is mainly due to the direct friends of the user getting a lot of weight. But, this is no problem in our applications, because for instance a recommendation system will not recommend the friends of the user anyway. In other words, we don't particularly care about the initial part of the vector.*

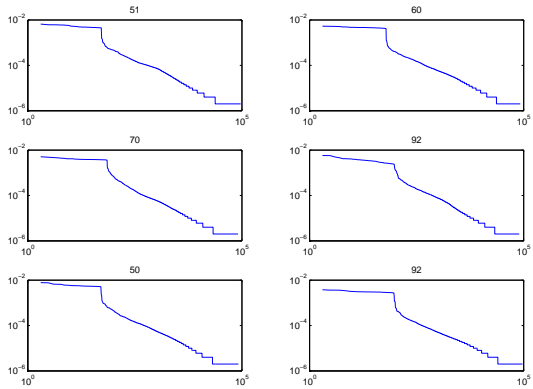

**Figure 3: Personalized PageRank power-laws for 6 random users; x-axis is index $i$ and y-axis is $i^{th}$ largest personalized PageRank. Both axes are log-scale.**

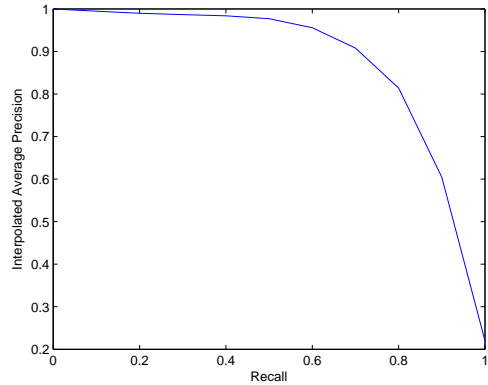

**Figure 5: 11 point interpolated average precision for top 1000 results.**

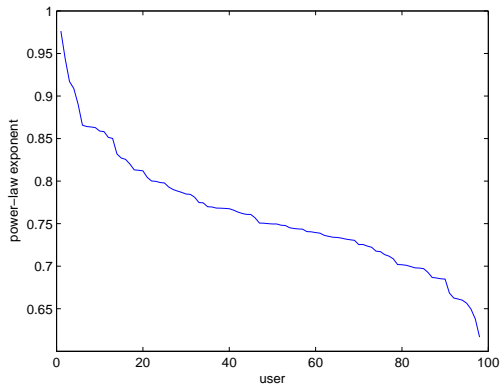

**Figure 4: Sorted power-law exponents for 100 random users.**

REMARK 4. *For the calculation of the power-law exponent of the personalized PageRank of each user (in figure 4), we only considered the part of the vector indexed between $[2f, 20f]$ where $f$ is the number of friends of the user. Again, the reason we did this was that this is really the only part of the vector which matters to us in our applications.*

### 4.4 A Few Random Steps Go a Long Way

As we showed in section 3.2, a relatively small number of random walk steps (e.g., as calculated in equation 4) are enough to approximately find the top $k$ personalized PageRank nodes. We experimentally tested this idea. To do so, notice that the stationary distribution of a random walk is the limit of the empirical walk distribution as the length of the walk approaches $\infty$. Thus, to accurately find the top $k$ nodes, we can theoretically do an infinite walk, and find the top $k$ most frequently visited nodes. Based on this idea, we did the following experiment: For each of the 100 randomly selected users, we did a random walk, personalized over that user, with 50000 steps. We calculated the top 100 most visited nodes, and considered them as the "true" top 100 results. Then, we did a 5000 step random walk for each user, and retrieved the top 1000 most visited nodes. For both experiments, we excluded nodes that were directly connected to the user. Then, we calculated the *11 point interpolated average precision* curve [34]. The result is given in figure 5. Notice that the curve validates our approach. For instance, the precision at recall level 0.8 is almost equal to 0.8, meaning that 80 of the top 100 "true" results were returned among the top 100 results of the short (5000 step) random walks. Similarly, precision of almost 0.9 at recall level 0.7 means 70 of the top 100 "true" results were retrieved among the top 77 results. This shows that even short random walks are good enough to find the top scoring nodes (in the personalized setting).

### 4.5 Number of Fetches

In theorem 8, we gave an upperbound on the number of fetches needed to compose a random walk out of the stored walk segments. We did an experiment to test this theoretical bound. In our experiments we found the average (over 100 users) number of fetches actually done to make a walk of length $s$, for $s$ between 100 and 50000, when we store $R$ walk segments per node, for each of the cases with $R \in \{5, 10, 20\}$. These are the thin lines in the plots in figure 6. We also calculated the corresponding theoretical upperbound on the number of fetches for each user (using its own power-law exponent), and then calculated the average over the 100 users. The results are the thick lines in the plots in figure 6. As can be seen in this figure, our theoretical bounds actually give an upperbound on the actual number of fetches in our experiments. Also, we see that the number of fetches that we make is not much sensitive to the number of stored random walks per node (i.e., $R$). Note that the theoretical guarantees are only valid for $R > q \ln n$ for a large enough constant $q$; hence the theoretical bound appears to be robust well before the range where we proved it.

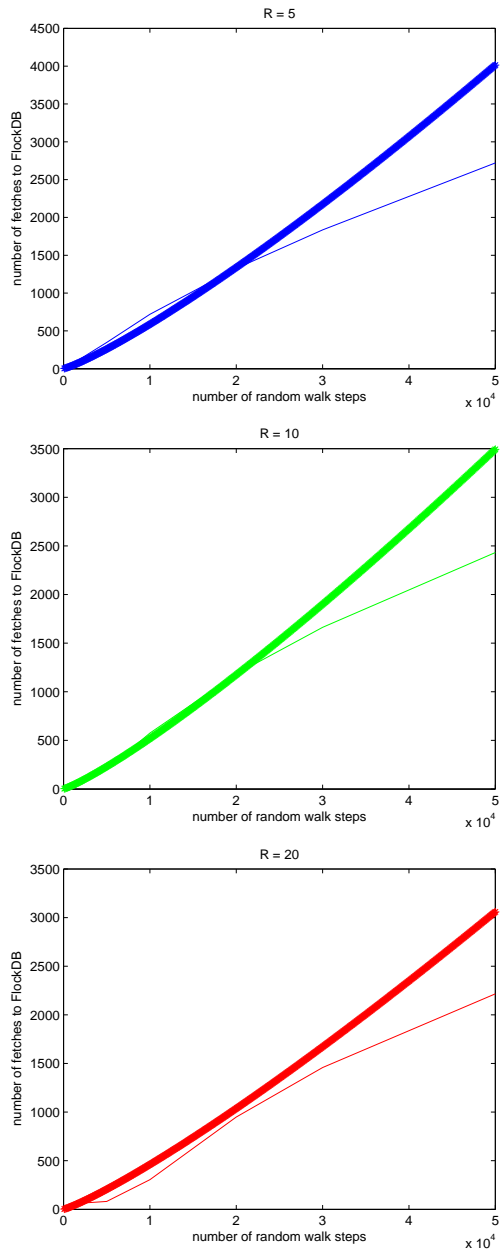

**Figure 6: Number of fetches (thin line: observed, thick line: theoretical bound).**

# APPENDIX

## A. EFFECTIVENESS OF RANDOM WALK BASED METHODS FOR LINK PREDICTION

As mentioned in the Introduction section, random walk based methods have been reported to be very effective for the link prediction problem on social networks [31]. We also did some experiments to explore this further. These are somewhat tangential to the rest of this paper, and by no means exhaustive. We present them not as significant original research but as an interesting data point for readers who are interested in practical aspects of recommendation systems.

|          | HITS | COSINE | PageRank | SALSA |
|----------|------|--------|----------|-------|
| Top 100  | 0.25 | 4.93   | 5.07     | 6.29  |
| Top 1000 | 0.86 | 11.69  | 12.71    | 13.58 |

Table 1: Link Prediction Effectiveness

We picked 100 random nodes from the Twitter social network. To select these users, we considered the network for two different dates, with 5 weeks of difference. Then, we selected random users who had a reasonable number of friends (between 20 and 30) on the first date, and increased the number of their friends by a factor between 50% and 100% by the second date. For the second date, we only counted the friends who already existed (and were reasonably followed, i.e., had at least 10 followers) on the first date (because, otherwise there is no way for a collaborative filter or link prediction system to find these users).

The reason we enforced the above criteria was that these users are among the typical reasonably active users on the network. Also, because they are increasing their friends set, they are good targets for a recommendation system.

For each of the 100 users, we used the network data from the first date to generate a personalized list of predicted links. We considered four link prediction methods: personalized PageRank, personalized SALSA, personalized HITS, and COSINE. We already explained personalized PageRank and personalized SALSA in the paper. Personalized HITS and COSINE also assign hub and authority scores to each node. For personalized HITS, when personalizing over node $u$, these scores are related as follows:

$$h_v = \epsilon \delta_{u,v} + (1-\epsilon) \sum_{\{x|\ (v,x)\in E\}} a_x$$

$$a_x = \sum_{\{v|\ (v,x)\in E\}} h_v$$

For the COSINE method, the hub score $h_v$ is defined as the cosine similarity of the neighbor sets of $u$ and $v$ (considered as 0-1 vectors). Then, the authority score, similar to HITS, is defined by:

$$a_x = \sum_{\{v|\ (v,x)\in E\}} h_v$$

We performed 10 iterations for each method to calculate the hub and authority scores. After generating the lists of predicted links, we calculated how many of the new friendships that were made by each user between the two dates were captured by the top 100 or top 1000 predictions. Finally, we averaged these numbers over the 100 selected users. The results are presented in table 1.

Notice that we do not expect these numbers to be large, because they are the number of friendships out of the prediction/recommendation list that the user made without being exposed to the recommendations at all. Also, in our experiments, each user had only 10-30 new friends, which is an upper bound on these numbers. This number would presumably be very different (i.e., much larger) if the user first received the recommendations and then decided which friendships to make. Nonetheless, the relative values of these numbers for different algorithms are good indicators of the predictive ability of those algorithms, specially when the differences are as pronounced as in our experiments. As we see from table 1, the systems based on random walks (i.e., Personalized PageRank and SALSA) perform the best: they significantly outperform HITS, and they also do better than the cosine similarity based link prediction system. These results are in accordance with the previous literature indicating the effectiveness of random walk based methods for the link prediction problem [31]. Moreover, it should be mentioned that there is also axiomatic support for this outcome [2, 6].

## B. PROOF OF THEOREM 1

It is already proved in [3] that $E[\widetilde{\pi}_v] = \pi_v$. So, we only need to prove the sharp concentration result. First, assume $R=1$. Fix an arbitrary node $v$. Define $X_u$ to be $\epsilon$ times the number of visits to $v$ in the walk stored at $u$, $Y_u$ to be the length of this walk, $W_u = \epsilon Y_u$, and $x_u = E[X_u]$. Then, $X_u$'s are independent, $\widetilde{\pi}_v = \frac{\sum_u X_u}{n}$ (hence $\pi_v = \frac{\sum_u x_u}{n}$), $0 \leq X_u \leq W_u$, and $E[W_u] = 1$. Then, it is easy to see that:

$$E[e^{tX_u}] \leq x_u E[e^{tW_u}] + 1 - x_u \leq e^{-x_u(1-E[e^{tW_u}])}$$

Thus:

$$Pr[\widetilde{\pi}_v \geq (1+\delta)\pi_v] \leq \frac{E[e^{tn\widetilde{\pi}_v}]}{e^{tn(1+\delta)\pi_v}}$$

$$= \frac{\prod_u E[e^{tX_u}]}{e^{tn(1+\delta)\pi_v}} \leq \frac{\prod_u e^{-x_u(1-E[e^{tW_u}])}}{e^{tn(1+\delta)\pi_v}}$$

$$= \frac{e^{-n\pi_v(1-E[e^{tW}])}}{e^{tn(1+\delta)\pi_v}} \leq e^{-n\pi_v \delta'}$$

where $W = \epsilon Y$ is a random variable with $Y$ having geometric distribution with parameter $\epsilon$, and $\delta'$ is a constant depending on $\delta$ (and $\epsilon$), found by an optimization over $t$.

Therefore, we see that if $\pi_v = \Omega(\ln n/n)$ (i.e., if $\pi_v$ is slightly larger than the average PageRank value $1/n$), then we already get a sharp concentration with $R=1$ (the analysis for $Pr[\widetilde{\pi}_v \leq (1-\delta)\pi_v]$ is similar, and hence we omit it here).

Now, assume we have $R$ walk segments stored at each node, where we do not necessarily have $R=1$. Then, similar to the above tail analysis, we get:

$$Pr[\widetilde{\pi}_v \geq (1+\delta)\pi_v] \leq e^{-nR\pi_v \delta'}$$

Therefore, choosing $R = \Omega(\frac{\ln n}{n\pi_v})$ we get exponentially decaying tails. Notice that this means even for average values of $\pi_v$ (i.e., for $\pi_v = \Theta(1/n)$), we have sharp concentration with $R$ as small as $O(\ln n)$.

This finishes the proof of the theorem.

## C. PROOF OF LEMMA 7

If $\widetilde{X}_{s,v}$ is the number of times that we visit $v$ when we take a random walk starting at the stationary distribution, then by coupling our walk with this stationary walk at the first reset time $t_s$ and all the steps afterwards, we can see that

$X_{s,v} - \widetilde{X}_{s,v} = X_{t_s,v} - \widetilde{X}_{t_s,v}$. Since $\sum_v X_{t_s,v} = \sum_v \widetilde{X}_{t_s,v} = t_s$, we get:

$$\sum_v |E[X_{s,v}] - s\pi_v| = \sum_v |E[X_{s,v}] - E[\widetilde{X}_{s,v}]|$$

$$= \sum |E[X_{t_s,v}] - E[\widetilde{X}_{t_s,v}]| \leq 2E[t_s] = \frac{2}{\epsilon}$$

which proves the lemma.